\documentclass{iauc}
\usepackage{graphicx}
\usepackage{subfigure}
\newcommand{\mnras}{{\it MNRAS\,\,}}
\newcommand{\apj}{{\it ApJ\,\,}}
\newcommand{\apjl}{{\it ApJ} (Letters)\,\,}
\newcommand{\araa}{{\it ARAA\,\,}}
\newcommand{\nat}{{\it Nature\,\,}}
\newcommand{\aj}{{\it AJ\,\,}}
\newcommand{\prd}{{\it PhysRevD\,\,}}
\newcommand{\prl}{{\it PhysRevLet\,\,}}
\newcommand{\aap}{{\it A\&A\,\,}}

\newcommand{\bc}{\begin{center}}
\newcommand{\be}{\begin{equation}}
\newcommand{\ee}{\end{equation}}
\newcommand{\ec}{\end{center}}

\newcommand{\lya}{Ly$\alpha$~}

\newcommand{\cfour}{\mbox{CIV}}
\newcommand{\sifour}{\mbox{SiIV}}
\newcommand{\osix}{\mbox{OVI}}
\newcommand{\ltsima}{\mbox{$\; \buildrel < \over \sim \;$}}
\def \simlt{\lower.5ex\hbox{\ltsima}}            
\def \gtsima{\mbox{$\; \buildrel > \over \sim \;$}}
\def \simgt{\lower.5ex\hbox{\gtsima}}            

\title[Conference summary: intergalactic medium] 
{ELT requirements for future observations of the Intergalactic Medium}

\author[Theuns \& Srianand]   
{Tom Theuns$^{1,2}$ and Raghunathan Srianand$^3$}%

\affiliation{
$^{1}$ Institute for Computational Cosmology, Department of Physics,
Durham University, UK\break
$^{2}$ Department of Physics, University of Antwerp, Belgium\break
and \break
$^{3}$ IUCAA, Post Bag 4, Ganeshkhind, Pune 411 007, India\break}

\begin{document}

\maketitle

\begin{abstract}
We summarise the science cases for an ELT that were presented in the
parallel session on the intergalactic medium, and the open discussion that
followed the formal presentations. Observations of the IGM with an ELT
provides tremendous potential for dramatic improvements in current
programmes in a very wide variety of subjects. These range from
fundamental physics (expansion of the Universe, nature of the dark
matter, variation of physical constants), cosmology (geometry of the
Universe, large-scale structure), reionisation (ionisation state of
the IGM at high $z\ge 6$), to more traditional astronomy, such as the
interactions between galaxies and the IGM (metal enrichment, galactic
winds and other forms of feedback), and the study of the interstellar
medium in high $z$ galaxies through molecules. The requirements on
ELTs and their instruments for fulfilling this potential are
discussed.  {} 

\end{abstract}

\firstsection 
\section{Introduction}
The advent of echelle spectrographs on 8m class telescopes since the
early 1990's has revolutionised our understanding of the intergalactic
medium (IGM) as observed in quasar spectra. These bright sources have
smooth intrinsic spectra with broad emission lines, yet the {\em
observed} spectra contain hundreds of narrow absorption lines due to
intervening absorbers. The latter can be studied in great detail from
the exquisite, $\rm S/N> 40$, spectra possible with UVES on VLT and HiRes
on Keck.

Most of the absorption in the UV is due to neutral hydrogen left over
from the Big Bang, forming a forest of lines (Bahcall \& Salpter 1965;
Gunn \& Peterson 1965; Lynds 1971). The weaker lines with column
density $N$({H~{\sc i})$\le 10^{15}{\rm cm}^{-2}$ are traditionally 
called the \lq
Lyman-$\alpha$ forest\rq, with the strongest lines with column density
$\ge 10^{20.3}{\rm cm}^{-2}$ that show a measurable damping-wing called
\lq Damped Lyman-$\alpha$ systems\rq\, (DLAs). The number of lines as a
function of redshift and column-density, $d^2N/dz/dN$(H~{\sc i}), is close
to a power-law $\propto~N$(H~{\sc i})$^{\beta}$ with $\beta<0$, as function
of column-density, and evolves strongly with redshift as fewer lines are
produced as the mean density decreases due to the expansion of the Universe,
see Rauch (1998) for a review.

The weaker lines originate in the filaments of the cosmic web which
itself is a natural outcome of how structure forms in a dark matter
dominated cosmology (Bi \etal\ 1992; Cen \etal\ 1994; Schaye
2001). The neutral hydrogen fraction is small at redshifts $\le 6$
(Gunn \& Peterson 1965) as the gas is photo-ionised and photo-heated by
the UV-background, with photo-ionisation rate $\Gamma$, produced by
galaxies and quasars (Haardt \& Madau 1996). At lower $z\le 2$, the
forest of lines thins-out into a Lyman-$\alpha$ \lq savanna\rq\,, but
the decline is slowed because  $\Gamma$ also decreases as the emissivity
from galaxies and QSOs drops (Theuns \etal\ 1998a; Dav\'e \etal\ 
1999). Conversely at increasing $z\ge 6$, the mean density increases,
but $\Gamma$ also decreases as many source have yet to form, turning
the forest into a Lyman-$\alpha$ \lq jungle\rq\, which absorbs (nearly)
all light, perhaps signaling the end of reionisation (Becker \etal\ 
2001; Djorgovski \etal\  2001). The forest provides a tremendous probe
of how the IGM evolves in the intermediate redshift range $2\le z\le
5$, because the absorbers are only mildly non-linear and hence can be
simulated reliably (Cen \etal\  1994; Hernquist \etal\  1996; Theuns \etal\  
1998b; Zhang \etal\  1998; Bryan \etal\  1999). The combination of
superb data with reliable models makes it possible to constrain models
and determine parameters.

Stronger lines form near galaxies, with the DLAs potential
proto-galaxies or proto-galactic lumps (Wolfe 1995; Haehnelt \etal\ 
1998; Ledoux \etal\ 1998).  Since these systems are discovered in absorption, it is a worry
that even denser systems might be missed because they make the
background QSO too faint to appear in a magnitude-limited survey. For a
recent appraisal of this issue see Ellison \etal\  (2005 and reference
therein).  DLAs shield the UV-background and some fraction of the gas
becomes molecular (see, e.g., Srianand this volume). The prospect of
studying star formation in small systems at high redshift which are too
faint to study in emission, is very exciting.

Quasar spectra also contain \lq metal\rq\, lines from highly ionised
species such as \cfour\,, \sifour\, and \osix\, (e.g., Cowie \etal\ 
1995). These metals were synthesised in stars and managed to diffuse
into the lower density surroundings, either as a result of galactic
winds, or due to an early generation of population~III~ stars.

The next section gives a short overview of recent results, with 
emphasis on opportunities for progress with the advent of new
observatories.

\section{IGM observations with ELT: science}
We begin with a short overview of numerical simulations of the IGM, as
these can be used to investigate the main limitations of current
observational strategies, thereby guiding the design for new
instruments. We then discuss current and future science that can be
done with IGM observations. The next section summarises the
corresponding requirements for an ELT.

\subsection{Hydrodynamical simulations}

\begin{figure}
\centering
  \includegraphics[height=6cm]{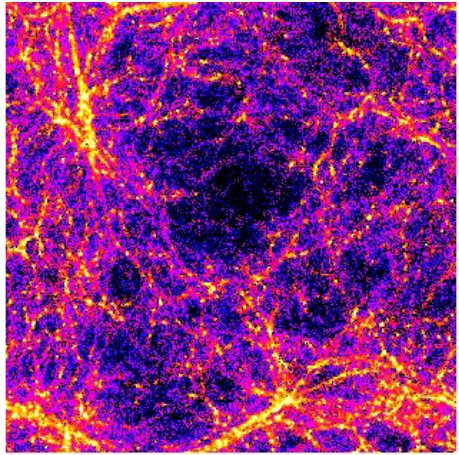}
  \caption{Baryon density in a slice through a 10Mpc/h simulation of a
  $\Lambda$CDM cosmology at redshift $z=3$. The gas traces the dark
  matter which follows a filamentary pattern called the \lq cosmic
  web\rq\,. Higher density halos at the intersection of filaments are
  the sites of galaxy formation. From Theuns \etal\ 
  1998b}\label{fig:ts_fig1}
\end{figure}

\begin{figure}
\centering
  \includegraphics[width=1.07\textwidth]{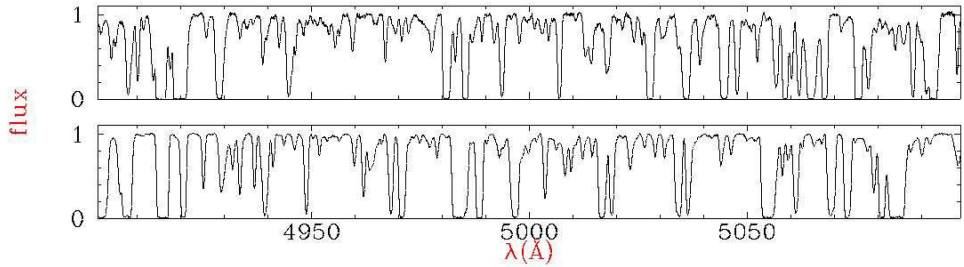}
  \caption{Comparison of a mock spectrum drawn from the simulation
  shown in Fig.~\ref{fig:ts_fig1} with a stretch of a HIRES spectrum
  of QSO\,1422+231 in the Lyman-$\alpha$ region. The mock spectrum has
  been smoothed to the instrumental resolution of HIRES, and noise has
  been added with similar dependence of S/N on flux and wavelength as
  the real data. Simulated and observed spectrum look very similar, in
  terms of the numbers and properties (such as width and strength) of
  absorption features they contain. This good correspondence means that
  it is possible to constrain some of the physical parameters that we
  know determine the line properties in the
  simulations.}\label{fig:ts_fig2}
\end{figure}

\begin{figure}
  \centering
  \subfigure[Simulation without feedback]{
          \label{fig:dl2858}
          \includegraphics[width=.45\textwidth]{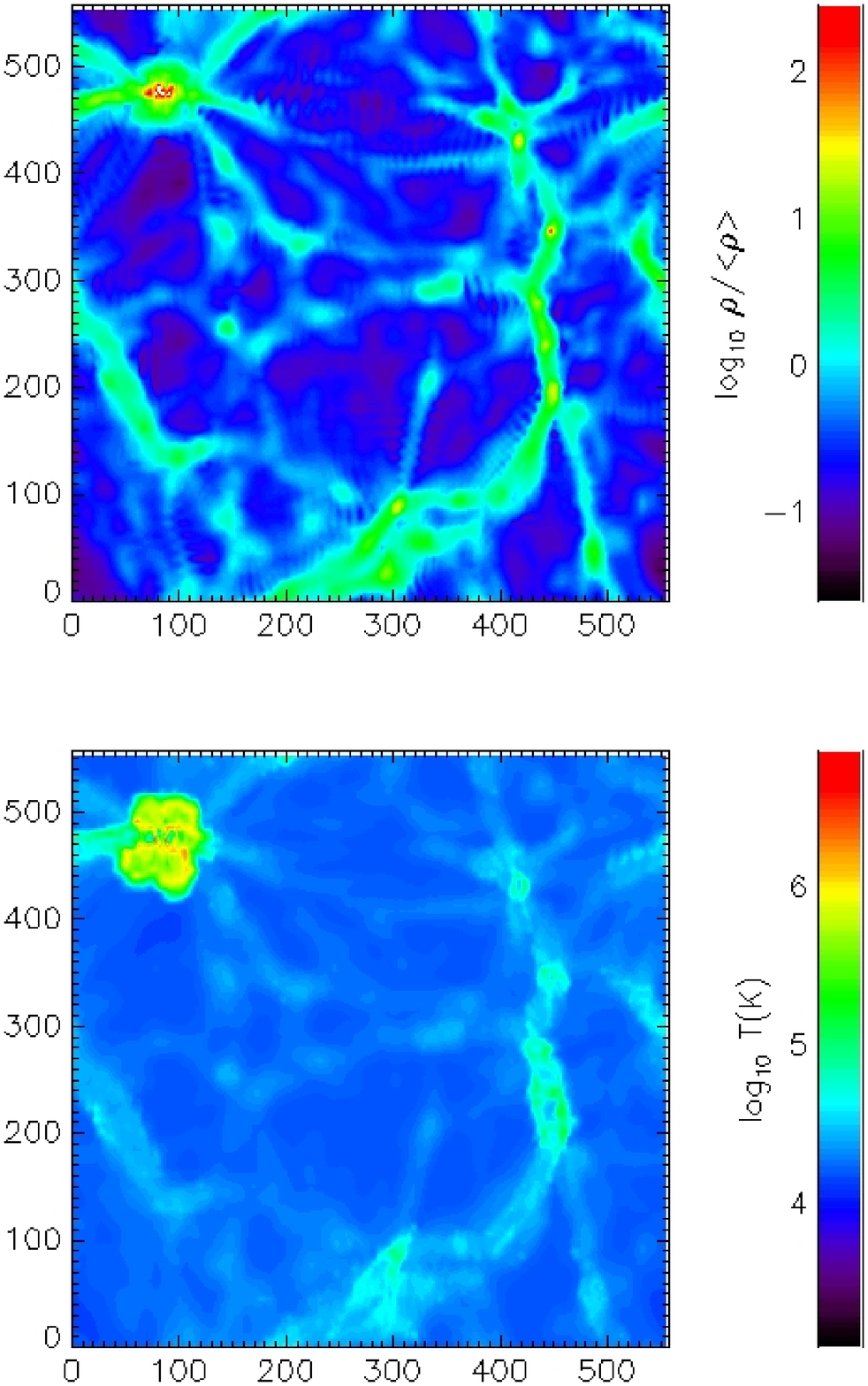}}
     \hspace{.3in}
     \subfigure[Simulation with feedback]{
          \label{fig:er2858}
          \includegraphics[width=.43\textwidth]{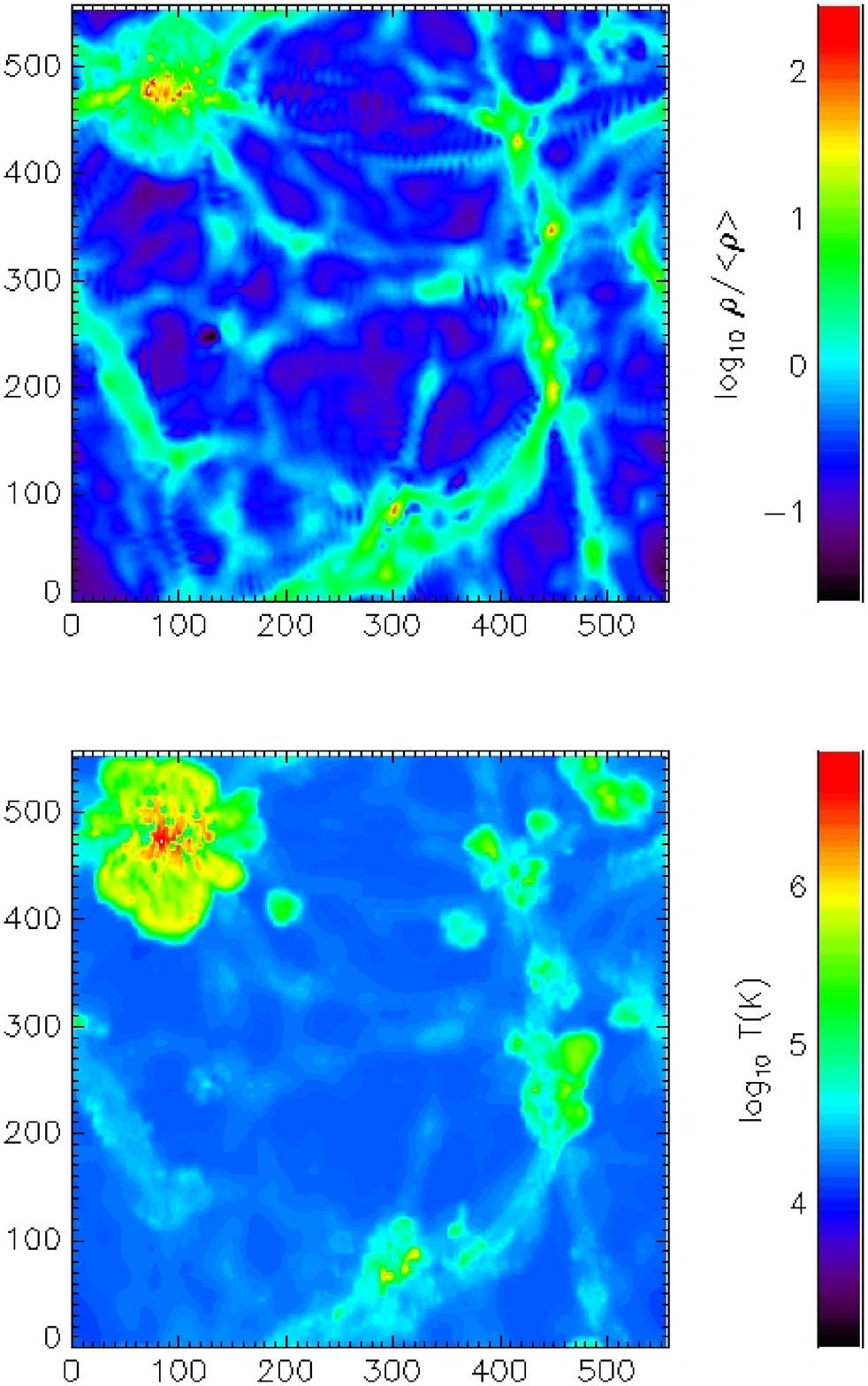}}
     \vspace{.3in}
  \caption{Density (top panels) and temperature (bottom panels) slices
  through a cosmological simulation (box size 5Mpc/h) at redshift $z=3$
  without (left panels) and with (right panels) feedback from star
  formation. Feedback disrupts many of the smaller galaxies and
  generates a hot bubble of shock heated and metal enriched gas around
  the bigger halos (for example in the top left corner) (From Theuns
  \etal\  2002a).
  }\label{fig:ts_fig3}
\end{figure}

Most of the weaker ``Lyman$-\alpha$ forest'' lines form in mildly over
dense or under dense structures that can be simulated accurately (Bi \etal\
  1992; Cen \etal\ 1994). Fig.~\ref{fig:ts_fig1} displays the gas
distribution in a cosmological hydrodynamical simulation at $z=3$, and
shows that the gas traces the filamentary pattern that results from
structure formation in a dark matter Universe. A sight line through
such a density distribution will most often go through low density
voids, occasionally intersecting a filament which will produce an
absorption line, and even more rarely pass close to or even straight
through a galaxy halo, producing a very strong absorption line.

Mock spectra generated from such simulations look very similar to the
real data, see, e.g., Fig.~\ref{fig:ts_fig2}. Since most of the lines are
due to structures that are only mildly over dense, it is possible to
simulate them quite reliably. Comparison of such simulated spectra with
observed ones makes it possible to constrain the model parameters and
investigate which cosmological parameters determine the line
statistics.

Most of the volume in the simulations is photo-heated by the
UV-background after reionisation, the volume affected by shocks from
structure formation is small. The temperature of the IGM affects the
properties of the lines (Theuns, Schaye \& Haehnelt 2000), because the
widths of the narrowest lines is restricted by thermal
broadening. Detailed comparison with data allows one to constrain the
thermal history of the IGM (Schaye \etal\  2000; Ricotti \etal\  2000;
Bryan \& Machacek 2000; McDonald \etal\  2001), because the thermal
time-scales are long in the low-density IGM. This also puts constraints
on reionisation if photo-heating is the dominant heating mechanism
(Theuns \etal\  2002b).

The detailed line properties are also sensitive to the nature of the
dark matter, and can for example constrain the mass of a putative warm
dark matter particle (Croft \etal\  1999; Viel \etal\  2005). If the
warm dark matter smoothing length is comparable to the width of
filaments, then this will affect the line shape. Note that these scales
become non-linear at lower $z$, making it much harder to put tight
constraints on the dark matter properties.

Sight lines passing close to a galaxy may be affected by
non-gravitational effects such as feedback from star formation or
AGN. Fig.~\ref{fig:ts_fig3} illustrates how supernova feedback causes
bigger galaxies to be embedded in a hot bubble of metal enriched gas,
which expands into the lower density surroundings of halos. Such sight
lines will also show metal line absorption, and it is possible to
compare in detail the metal-hydrogen correlation in simulations and
data to infer the physical properties in the surroundings of high-$z$
galaxies (Adelberger \etal\  2003; Pieri \etal\  2005). ELTs will
provide dramatic improvements in this subject, because the bigger
collecting area will allow one to observe fainter sources, and hence
allow a far finer grid of sight lines probing galaxy environs.

\subsection{Fundamental Physics}
High resolution high signal-to-noise spectra of high-$z$ QSOs are 
frequently used to test the current theories of cosmology and 
fundamental physics.\\
\noindent {\em CODEX:} The observed wavelength of a given absorption line changes
with time due to the expansion of the Universe. The COsmic Dynamics 
EXperiment (CODEX) (see Molaro \etal\ , this volume) aims to measure 
this change directly by observing many lines in a QSO spectrum at 
extreme signal-to-noise and resolution, and repeat the measurement a 
few decades later. Such an experiment therefore also requires very 
high and long term stability of the spectrograph.

A more indirect measure of the expansion of the Universe is by
determining the CMB temperature at intermediate redshift using the
fine-structure excitation lines of carbon in DLAs, which is excited by
CMB photons (see \cite{srianand00}). Detecting C~{\sc i} absorption
lines from the low density regions, where the collisional excitation
will be sub-dominant using high S/N spectra will allow one to directly
map the redshift evolution of temperature of the CMBR (R. Srianand,
this volume).

Some of the current theories of fundamental physics, such as SUSY, GUT
and Super-string theory, allow possible space and time variations of
the fundamental constants. QSO absorption lines can be used to probe
the time evolution of fundamental constants. The heavy element
absorption lines and H$_2$ Lyman Werner band absorptions lines are used
to investigate the time-variation of the electromagnetic coupling
constant $\alpha$ (see \cite{murphy03}; \cite{chand04} and Mollaro \etal\
 this volume) and the proton to electron mass ratio ($\mu$)(see
\cite{ivanchik05}), respectively. The available constraints based on 8m
class telescopes are still much higher than those achieved by
terrestrial techniques. Higher resolution (R$\ge100\,000$) and good
signal-to-noise ratios ($>100$) are needed to improve the precession.

The Square Kilometer Array (SKA) will measure the 21-cm line in most of
the DLAs. The wavelength of the transition depends on $\alpha$, $\mu$
and the proton g-factor and can provide a combined constraint on the
variation of all these fundamental constants (see
\cite{curran04}). Detecting H$_2$ and weak transitions of Mn~{\sc ii},
Ni~{\sc ii} in DLAs with high signal-to-noise and resolution will allow
us to lift the degeneracy between the variation of different fundamental
constants that decide the shift of the 21\,cm absorption line.

To avoid the systematics caused by the small-scale properties of the
lines we require high resolution and high signal-to-noise to improve
current constraints. To study the redshift evolution and to be able to
use different sets of lines from the same system, it is of paramount
importance to have a wide wavelength coverage.

\subsection{Cosmology}
The large-scale flux distribution can be used to infer the dark matter
power spectrum (Croft \etal\  1998; McDonald \etal\  2000; Viel \etal\ 
2004) and constrain the neutrino mass (Croft \etal\  1999; Viel \etal\ 
2005). These measurements are currently limited by uncertainties in the
shape of the QSO's underlying continuum, calibration of the echelle
spectrograph for high resolution spectra, and by the statistics of
available spectra, and would not obviously benefit from an
ELT. However, the influence of large-scale structure is also very
prominent in simulations at {\em low} optical depths, below the median
$\tau$. Such observations require much higher S/N than currently
available, since even S/N=50 data cannot recover the median optical
depth at $z\sim 2$.

Observations of the forest along parallel sight lines can constrain the
topology of the Universe via the Alcock-Paczynski (1979) test
(e.g., Rollinde \etal\  2003). Such a project would benefit greatly from
observing fainter QSOs and bright LBGs to improve the sampling in the
transverse direction to the line of sight.

\subsection{Reionisation}

The Lyman-$\alpha$ forest becomes increasingly opaque above $z\ge 6$,
perhaps signaling the end of reionisation. If the IGM is polluted
through winds from early generations of dwarf galaxies then the
ionisation state of the IGM can be probed through the absorption
produced by C~{\sc iv}, C~{\sc ii}, O~{\sc i} and Si~{\sc ii} (with NIR
wavelength $\lambda<2.1\mu$m for $z<12.5$, 14.7, 15.1 and 15.7,
respectively, e.g., Oh 2002). Given the rapidly declining space density
of QSOs, Gamma Ray Bursts (GRBs) or super novae could be used as
background sources.

GRBs have mean afterglow fluxes of 1.5 to 0.05$\mu$J at $z=10$, 1 to 10
days after the explosion ($K_{AB}=23.6$ to 27). High resolution
($R=4\times 10^4$) and high signal-to-noise ratios ($>50$) are required to
detect individual lines in the NIR. Observations of a very bright
$4\mu$J GRB at $R=10^4$ and S/N=50 gives detection limits for $N$({\rm
C~{\sc ii}})=$4\times 10^{12}{\rm cm}^{-2}$ and $N$(O~{\sc
i})=1$\times 10^{13}{\rm cm}^{-2}$ . Pair-instability SNe of
$M=140-260M_\odot$ pop.~III precursors have $K_{\rm AB}=25$ for
$z=10-15$, and are also potential targets, with a possible time-lag of
weeks between discovery and ELT follow-up spectroscopy (see the
presentation by J. Bergeron in this volume for details).

\subsection{Galaxy-Intergalactic medium interactions and metal enrichment}
The metal density of carbon as inferred from C~{\sc iv} pixel optical
depth analysis (Songaila 2001; Schaye \etal\  2003) shows little
evidence for evolution over the redshift range $z=2-5$, with possibly a
decline by factor of two above $z=6$. It is possible that not all
metals are seen. Just as most metals are in the hot intra-cluster gas
at $z<1$, metals could be in hot gas resulting from galactic winds at
higher $z$, thereby not producing significant C~{\sc iv} absorption
(Theuns \etal\  2002a and this volume). The shape of the UV-background,
and its evolution with $z$, is the main uncertainty in converting
optical depth to metallicity. Improved constraints require the
detection of many more transitions to eliminate this uncertainty.

What is the origin of the metals seen in the IGM? Are the metals due to
galactic winds, or is some fraction the result of pop.~III stars?  This
important question can be addressed by correlating metals seen in
absorption with the presence of galaxies (e.g., Adelberger \etal\  2003,
2005; Pieri, Schaye \& Aguirre 2005). This can be done by probing the
IGM with many sight lines, and requires obtaining spectra of fainter
sources, including the brighter Lyman-break galaxies (LBGs)
themselves. Current state of the art (Adelberger \etal\  2005) is
limited to measuring the mean metallicity in C~{\sc iv} as function of
the galaxy's impact parameter; higher S/N should make it possible to
look for metals in each individual galaxy spectrum and obtain good
redshifts.
A better understanding of galaxy-IGM interactions is needed to
constrain how feedback from stars and AGN affects galaxy formation as
a function of redshift and galaxy mass. The redshift range $z\sim 3$
is well suited for such a study, as there are many lines in the
observed optical-NIR part of the spectrum suited to ground-based
observations, but an ELT is required to be able to observe fainter
QSOs or brighter LBGs and dramatically improve the sampling with many
more sight lines.

\subsection{Molecules at high $z$}

Detecting H$_2$ and other molecules at high-z through their electronic
transitions is important for understanding the physical conditions and
astrochemistry in the interstellar medium of galaxies and
protogalaxies at a very early epoch.
Up to now, H$_2$ has been detected in $\sim15\%$ of DLAs
(\cite{ledoux03}) and only one system shows detectable HD
(Varshalovich \etal\ 2001). H$_2$ can
potentially be detected from LBGs and GRB host galaxies. This will
allow us to understand the interstellar medium in these early
galaxies.  As the Lyman Werner band absorptions of H$_2$ are
expected in the \lya forest, it is important to have high resolution R
= 20\,000 and signal-to-noise ($>20$) in the blue spectrum. ELTs with
blue sensitive spectrographs can allow us to search for H$_2$ in
fainter QSOs, GRBs and brighter LBGs. However, in the case of GRBs,
H$_2$ may be in non-equilibrium and it is important to target the
source as quickly as possible to be able to detect the H$_2$ lines and
follow the time variation of H$_2$ column density. This will give
important clues about the GRB hosts.

Detecting other molecules in systems with H$_2$ is also important for
the understanding of astrochemistry in low metallicity gas in the
early universe.  CO is not detected in DLAs and the achieved limits
are close to the lowest column measured in Milky Way. As the
metallicities in DLAs are low, to test N(H$_2$) vs. N(CO) relation we
need to push this limit by roughly a factor 50 (see the presentation
by R. Srianand in this volume).

\section{IGM observations with ELT: requirements}

In this section we summarise the ELT requirements that emerge out of
the discussion in the parallel session on IGM.

\noindent{\em {High $z>7$}}: {\bf Reionisation, metals, Lyman-$\alpha$
emitters:} Constraining IGM enrichment and its ionisation state from
metal lines at $z>7$ requires observations at intermediate resolution
of $R=2000$ in the NIR with S/N up to 100. An OH-line suppressor with
multiple IFUs with field-of-view of several arcmin$^2$ is ideal.
Targets are moderately faint QSOs and Lyman-break galaxies of
$m_{AB}\sim 27$, but require an ELT larger than 30m. NIR observations
at $R=10^4$ with S/N up to 100 is possible from average-luminosity
GRBs and pop.~III SNe. Targets need to be found using dedicated
ground and space-based telescopes.
\begin{enumerate}
\item[$\bullet$] NIR, R=2000, S/N=100 (bright QSOs)
\item[$\bullet$] NIR, R=10000, S/N=100 (single target GRBs)
\end{enumerate}

\noindent{\em {Intermediate $z<7$}: {\bf Metals:}} High resolution 
$R=40000$ and S/N of 10000 optical spectra of bright QSOs 
($z=2-5$, $m_{AB}=16-17$) and bright GRBs ($z$ up to 7, lag is 1 day, 
$m_{AB}=20$, S/N=100) in single target mode are required to study the 
distribution of metals in the IGM, and its evolution with $z$. 
The spectrograph should be {\em blue sensitive} and have a large 
wavelength range ($\lambda=3030-9300\AA$) to be able to cover a 
large range of transitions and constrain the ionisation
corrections. The latter is the major uncertainty in inferring
metallicity, so a large $\lambda$ range is essential.
\begin{enumerate}
\item[$\bullet$] optical, R=40000, S/N=$10^3-10^4$ (single target
  bright QSOs, GRBs). Blue sensitive, large $\lambda$ ($3030-9300\AA$)
  coverage.
\end{enumerate}

\noindent{\em {Lower $z<5$}: {\bf galaxy-IGM connection, UV-escape
from galaxies:}} The main gain of an ELT is the possibility of
observing fainter QSOs, which allows one to sample the metal
distribution in the IGM, and its correlation with galaxies
dramatically better by providing a much finer grid of lines along
which the IGM can be probed. The QSOs and bright LBGs can be observed
with optical, high $R=50000$ spectroscopy (S/N=100) to probe the
distribution of metals. A detailed correlation of these metals with
galaxies requires the redshift determination of the fainter LBGs (up
to 0.01$L_\star$) using optical/NIR MOS of $R=2000-5000$, with
multiple IFUs with a total FoV of several arcmin$^2$, centered on LBGs
and QSOs. NIR is required to obtain good redshifts for the galaxies
from stellar and ISM lines, since many of the UV-lines can be
significantly off-set from the redshift of the stars.
\begin{enumerate}
\item[$\bullet$] Optical/NIR, R=2000-5000, S/N=100 (0.01$L_\star$~LBGs)
  with multiple IFUs, FoV several arcmin$^2$
\item[$\bullet$] Optical, R=50000, S/N=100 (bright LBGs, QSOs)
\end{enumerate}

Many small programmes could be started at early stages of construction
if the instruments are available. 8m-class telescopes will be used to
find (candidate) LBGs and Lyman-$\alpha$ emitters.

\begin{acknowledgments}
We wish to thank IAU for a travel grant. TT thanks PPARC for the award
of an Advanced Fellowship, and J Schaye, R Bower and I Smail for
comments on the draft.
\end{acknowledgments}


\begin{thebibliography}{}

\bibitem[Adelberger \etal\ 2003]{adelberger03} Adelberger, K.L., Steidel,
C.C., Shapley, A.E. \& Pettini, M. 2003, \apj 584, 45

\bibitem[Adelberger \etal\ (2005)]{2005ApJ...629..636A} Adelberger, K.L., 
Shapley, A.E., Steidel, C.C., Pettini, M., Erb, D.K. \& Reddy, N.A. 
2005, \apj 629, 636 


\bibitem[Alcock \& Paczynski(1979)]{1979Natur.281..358A} Alcock, C. \& 
Paczynski, B. 1979, \nat 281, 358 

\bibitem[Bahcall \& Salpeter(1965)]{1965ApJ...142.1677B} Bahcall, J.~N., \& 
Salpeter, E.~E.\ 1965, \apj 142, 1677 

\bibitem[Becker \etal\ (2001)]{2001AJ....122.2850B} Becker, R.~H., 
Fan, X., White, R., \etal\  2001, \aj 122, 2850 


\bibitem[Bi \etal\ (1992)]{1992A&A...266....1B} Bi, H.~G., Boerner, G., \& 
Chu, Y.\ 1992, \aap 266, 1 


\bibitem[Bryan \& Machacek(2000)]{2000ApJ...534...57B} Bryan, G.~L., \& 
Machacek, M.~E.\ 2000, \apj 534, 57 

\bibitem[Bryan, Machacek, Anninos, \&
Norman(1999)]{1999ApJ...517...13B} Bryan, G.~L., Machacek, M.,
Anninos, P., \& Norman, M.~L.\ 1999, \apj 517, 13



\bibitem[\protect\citename{Cen \etal 1994}]{cenor94} Cen, R.,
Miralda'Escud\'e, J., Ostriker, J.P.\&  Rauch, M., 1994, \apjl 437, L83

\bibitem[Chand \etal\ 2004]{chand04} Chand, H., Srianand, R.,
  Petitjean, P. \etal\ 2004, \aap 417, 853

\bibitem[Curran \etal\ 2004]{curran04} 	Curran, S. J., Kanekar, N \&
Darling, J. K., 2004, {\it New Astronomy Reviews} 48, 1095


\bibitem[Cowie, Songaila, Kim, \& Hu(1995)]{1995AJ....109.1522C} Cowie, 
L.~L., Songaila, A., Kim, T., \& Hu, E.~M.\ 1995, \aj 109, 1522 

\bibitem[Croft \etal\ 1998]{croft98} Croft, R. A. C. Weinberg, D. H. Katz, N., \& Hernquist, L, 1998, \apj  495, 44


\bibitem[Croft \etal\ (1999)]{1999PhRvL..83.1092C} Croft, R.~A.~C., Hu, W., 
\& Dav{\'e}, R.\ 1999, \prl 83, 1092 


\bibitem[Dav{\'e}, Hernquist, Katz {\&} Weinberg
(1999)]{1999ApJ...511..521D} Dav{\'e}, R., Hernquist, L., Katz, N.,
\& Weinberg, D. 1999, \apj 511, 521

\bibitem[Djorgovski \etal\ (2001)]{2001ApJ...560L...5D} Djorgovski, S.~G., 
Castro, S., Stern, D., \& Mahabal, A.~A.\ 2001, \apjl 560, L5 

\bibitem[Ellison \etal\ (2005)]{2005AJ....130.1345E} Ellison, S.~L., Hall, 
P.~B., \& Lira, P.\ 2005, \aj 130, 1345 


\bibitem[Gunn \& Peterson(1965)]{1965ApJ...142.1633G} Gunn, J.~E., \& 
Peterson, B.~A.\ 1965, \apj 142, 1633 

\bibitem[Haardt \& Madau(1996)]{1996ApJ...461...20H} Haardt, F., \& Madau, 
P.\ 1996, \apj 461, 20 

\bibitem[Haehnelt \etal\ (1998)]{1998ApJ...495..647H} Haehnelt, M.~G., 
Steinmetz, M., \& Rauch, M.\ 1998, \apj 495, 647 


\bibitem[Hernquist \etal\ (1996)]{1996ApJ...457L..51H} Hernquist, L., Katz, 
N., Weinberg, D.~H., \& Miralda-Escud{\'e}, J.\ 1996, \apjl 457, L51 

\bibitem[Ivanchik \etal\ 2005]{ivanchik05}{ Ivanchik. A., Petitjean, P., 
Varshalovich, D., Aracil, B., Srianand, R., Chand, H., Ledoux, C. \& 
Boisse, P. 2005, \aap 440, 45}
%
\bibitem[Ledoux et al.(1998)]{1998A&A...337...51L} Ledoux, C., Petitjean, 
P., Bergeron, J., Wampler, E.~J., \& Srianand, R.\ 1998, \aap 337, 51 

\bibitem[Ledoux \etal\ 2003]{ledoux03} Ledoux, C., Petitjean, P., \& Srianand, R. 2003, \mnras 346, 209

\bibitem[Lynds(1971)]{1971ApJ...164L..73L} Lynds, R.\ 1971, \apjl 164,
  L73 

\bibitem[McDonald \etal\ 2000]{mcdonald00} McDonald P., Miralda-Escude, J., 
Rauch, M., Sargent W. L. W., Barlow, T. A., Cen, R \& Ostriker, J. P. 2000, 
\apj 543, 1 

\bibitem[McDonald \etal\ (2001)]{2001ApJ...562...52M} McDonald, P., 
Miralda-Escud{\'e}, J., Rauch, M., Sargent, W.~L.~W., Barlow, T.~A., \& 
Cen, R.\ 2001, \apj 562, 52 

\bibitem[Murphy \etal\ 2003]{murphy03}  Murphy, M. T., Webb, J. K. \&  Flambaum, V. V. 2003, \mnras 345, 609 

\bibitem[Oh(2002)]{2002MNRAS.336.1021O} Oh, S.~P.\ 2002, \mnras 336, 1021 

\bibitem[]{} Pieri, M.M., Schaye, J. \&  Aguirre, A., 2005, preprint, astro-ph/0507081

\bibitem[Rauch(1998)]{1998ARA&A..36..267R} Rauch, M.\ 1998, \araa 36,
  267 

\bibitem[Ricotti \etal\ (2000)]{2000ApJ...534...41R} Ricotti, M., Gnedin, 
N.~Y., \& Shull, J.~M.\ 2000, \apj 534, 41 

\bibitem[]{} Rollinde, E., Petitjean, P., Pichon, C., Colombi, S., 
Aracil, B., D'Odorico, V. \&  Haehnelt, M. G. 2003, \mnras 341, 1279. 

\bibitem[Schaye \etal\ (2000)]{2000MNRAS.318..817S} Schaye, J., Theuns, T., 
Rauch, M., Efstathiou, G., \& Sargent, W.~L.~W.\ 2000, \mnras 318, 817 

\bibitem[Schaye(2001)]{2001ApJ...559..507S} Schaye, J.\ 2001, \apj 559, 
57

\bibitem[Schaye \etal\ (2003)]{2003ApJ...596..768S} Schaye, J., Aguirre, A., 
Kim, T., Theuns, T., Rauch, M., \& Sargent, W.~L.~W.\ 2003, \apj 596, 768 

\bibitem[Songaila(2001)]{2001ApJ...561L.153S} Songaila, A.\ 2001, \apjl
561, L153 

\bibitem[Srianand, Petitjean \& Ledoux 2000]{srianand00} Srianand R.,
  Petitjean P. \&  Ledoux C. 2000, \nat 408, 931

\bibitem[Theuns \etal\ (1998)]{1998MNRAS.297L..49T} Theuns, T., Leonard, A., 
\& Efstathiou, G.\ 1998a, \mnras 297, L49 

\bibitem[Theuns \etal\ (1998)]{1998MNRAS.301..478T} Theuns, T., Leonard, A., 
Efstathiou, G., Pearce, F.~R., \& Thomas, P.~A.\ 1998b, \mnras 301, 478 

\bibitem[Theuns \etal\ (2000)]{2000MNRAS.315..600T} Theuns, T., Schaye, J., 
\& Haehnelt, M.~G.\ 2000, \mnras 315, 600 


\bibitem[Theuns \etal\ (2002)]{2002ApJ...578L...5T} Theuns, T., Viel, M., 
Kay, S., Schaye, J., Carswell, R.~F., \& Tzanavaris, P.\ 2002a, \apjl 578, 
L5 

\bibitem[Theuns \etal\ (2002)]{2002ApJ...567L.103T} Theuns, T., Schaye, J., 
Zaroubi, S., Kim, T.-S., Tzanavaris, P., \& Carswell, B.\ 2002b, \apjl 567, 
L103 

\bibitem[]{} Varshalovich, D. A., Ivanchik, A. V., Petitjean, P.,
  Srianand, R. \& Ledoux, C. 2001, {\it AstL} 27, 683

\bibitem[Viel \etal\ (2004)]{2004MNRAS.354..684V} Viel, M., Haehnelt, M.~G., 
\& Springel, V.\ 2004, \mnras 354, 684 

\bibitem[Viel \etal\ (2005)]{2005PhRvD..71f3534V} Viel, M., Lesgourgues, J., 
Haehnelt, M.~G., Matarrese, S., \& Riotto, A.\ 2005, \prd 71, 063534 

\bibitem[Wolfe 1995]{wolfe95} Wolfe A. M. 1995, in: Meylan G. (ed.),
  Proc. ESO Workshop, {\it QSO Absorption lines} (Berlin: Springer), p. 13

\bibitem[Zhang \etal\ (1998)]{1998ApJ...495...63Z} Zhang, Y., Meiksin, A., 
Anninos, P., \& Norman, M.~L.\ 1998, \apj 495, 63 

\end{thebibliography}
\end{document}